\documentclass[10pt,twocolumn,showpacs,preprintnumbers,amsmath,amssymb,aps,prl,superscriptaddress,longbibliography]{revtex4-2}
\usepackage{appendix}
\usepackage{bbm}
\usepackage{mathrsfs}
\usepackage{graphicx}
\usepackage{dcolumn}
\usepackage{bm}
\usepackage{braket}
\usepackage{amsmath}
\usepackage{amsfonts}
\usepackage[dvipsnames]{xcolor}
\usepackage[colorlinks=true,linkcolor=Blue,urlcolor=BlueViolet,citecolor=BlueViolet]{hyperref}
\usepackage{natbib}
\usepackage{floatrow}

\begin{document}

\title{FeTa$X_2$: A ferrimagnetic quantum anomalous Hall insulator}
\author{Yadong Jiang}
\affiliation{State Key Laboratory of Surface Physics and Department of Physics, Fudan University, Shanghai 200433, China}
\affiliation{Shanghai Research Center for Quantum Sciences, Shanghai 201315, China}
\author{Huan Wang}
\affiliation{State Key Laboratory of Surface Physics and Department of Physics, Fudan University, Shanghai 200433, China}
\affiliation{Shanghai Research Center for Quantum Sciences, Shanghai 201315, China}
\author{Jing Wang}
\thanks{Contact author: wjingphys@fudan.edu.cn}
\affiliation{State Key Laboratory of Surface Physics and Department of Physics, Fudan University, Shanghai 200433, China}
\affiliation{Shanghai Research Center for Quantum Sciences, Shanghai 201315, China}
\affiliation{Institute for Nanoelectronic Devices and Quantum Computing, Fudan University, Shanghai 200433, China}
\affiliation{Hefei National Laboratory, Hefei 230088, China}

\begin{abstract}
We propose that the van der Waals layered ternary transition metal chalcogenides FeTa$X_2$ ($X=$ S, Se, Te) as a new family of ferrimagnetic quantum anomalous Hall insulators with a sizable bulk gap and high Chern number $\mathcal{C}=-2$. The magnetic order arises primarily from Fe atoms, whose strong ferromagnetic exchange induces moments on neighboring Ta sites. The large topological gap originates from a \emph{deep} $s$–$d$-type band inversion between spin-down Ta $d_{z^2}$ and $d_{xy}$ orbitals near the Fermi level—a mechanism unique to $d$-orbital systems. Remarkably, the Curie temperature of monolayer FeTa$X_2$ is predicted to significantly exceed that of monolayer MnBi$_2$Te$_4$. Furthermore, both the Curie temperature and topological gap scale positively with the spin-orbit coupling strength of the $d$ electrons, suggesting a common physical origin. Owing to its structural similarity to Fe$X$ superconductors and reduced net magnetization, FeTa$X_2$ further offers a promising platform for realizing chiral topological superconductivity. These findings, if realized experimentally, could open new avenues for the research and application of topological quantum physics.
\end{abstract}


\maketitle

\emph{Introduction---}The discovery of the quantum anomalous Hall (QAH) effect marked a milestone in the study of topological quantum states~\cite{tokura2019,chang2023,wang2017c,bernevig2022}. These states exhibit an insulating bulk characterized by a finite Chern number and support gapless chiral edge modes~\cite{thouless1982,haldane1988,halperin1982}. The QAH phase provides a versatile platform for realizing exotic phenomena such as Majorana edge modes~\cite{fu2009a,akhmerov2009,qi2010b,wang2015c,lian2018b} and the quantized topological magnetoelectric effect~\cite{qi2008,essin2009,morimoto2015,wang2015b,mogi2017}, with potential for dissipationless electronic applications~\cite{zhang2012,wang2013a}. Experimental realizations of the QAH effect span a range of material systems~\cite{chang2013,chang2015,mogi2015,bestwick2015,watanabe2019,deng2020,li2024,lian2025,guo2025quantized,Serlin2020,Li2021,park2023,xu2023,lu2024,han2024,sha2024}, yet all occur at liquid helium temperatures. This low critical temperature remains a major barrier to practical use, including in metrological standards~\cite{okazaki2022,patel2024,huang2025}. Identifying new QAH insulators~\cite{you2019,sunj2020,sun2020,liy2020,li2022,xuan2022,jiang2023,yao2024,jiang2024} with higher transition temperatures and larger bulk gaps is therefore a central challenge in topological materials research. 

The QAH effect originates from spin-polarized band inversion~\cite{liu2008}, where achieving high transition temperatures and large gaps requires a delicate interplay between two seemingly conflicting requirements: strong spin-orbit coupling (SOC)—favored by heavy elements and stable ferromagnetism—typically emerges in metallic systems composed of lighter $3d$ element. This incompatibility has not been resolved in existing QAH systems. In magnetic topological insulators (TIs), the inhomogeneities drastically suppress the exchange gap by several orders of magnitude~\cite{yu2010,zhang2019,li2019,otrokov2019a,chong2020,garnica2022}; moir\'e systems are intrinsically limited by their low energy scales~\cite{Li2021,park2023,xu2023}; and rhombohedral graphene suffers from a weak proximitized Ising SOC~\cite{liu2025}. Therefore, finding stoichiometric 2D magnetic materials for the QAH effect—ideally in monolayer form with versatile tunability—is highly desired.

\begin{table}[b]
\caption{Lattice constant $a$; Curie temperature $T_c$ from Monte Carlo simulations; band gap $E_g$; magnetocrystalline anisotropy energy (MAE) per unit cell $E_{\rm{MAE}}$, defined as the total energy difference between in-plane and out-of-plane spin configurations.}
\begin{center}\label{tab1}
\renewcommand{\arraystretch}{1.4}
\begin{tabular*}{\columnwidth}
{@{\extracolsep{\fill}}ccccc}
\hline
\hline
Materials &$a$~(\r{A})  & $T_c$~(K) & $E_g$~(meV)  & $E_{\rm{MAE}}$~(meV)\\
\hline
FeTaS$_2$  & 3.74 & 725 & 363.3 & 11.2 \\
FeTaSe$_2$ & 3.80 & 770 & 401.7 & 13.0 \\
FeTaTe$_2$ & 3.91 & 560 & 218.7 & 18.5 \\
\hline
\hline
\end{tabular*}
\end{center}
\end{table}

A promising route toward high-temperature QAH insulators lies in leveraging correlated $d$ electrons, which can simultaneously support robust magnetism and nontrivial band topology. FeSe and related compounds exemplify this potential, as electron doping via lithium intercalation has revealed the coexistence of superconductivity, itinerant ferromagnetism, and topology~\cite{lei2017,ying2018,kim2023,hu2025Li,xu2016,Wang2018science,Zhangpeng2018science}. Motivated by these advances, we perform density functional theory (DFT) calculations and tight-binding modeling to show that monolayer FeTa$X_2$ ($X$ = S, Se, Te) is a compelling candidate for high-temperature QAH insulators. 
Structural relaxations and magnetic properties are obtained within $\text{DFT}+\text{Hubbard}$ $U$~\cite{kresse1996,dudarev1998,perdew1996}, while the electronic structure and MAE are confirmed using the Heyd-Scuseria-Ernzerhof hybrid functional~\cite{hse06}. All members of the FeTa$X_2$ family exhibit a ferrimagnetic ground state with high Chern number $\mathcal{C}=-2$ and exceptionally large topological gaps. The gap originates from a deep $s$-$d$ band inversion between Ta $d_{z^2}$ and $d_{xy}$ orbitals, while the strong ferrimagnetic order is driven by Fe atoms. This cooperative mechanism—where both magnetism and topology emerge from $d$-orbital physics—highlights a general design principle for engineering robust QAH phases in correlated materials.

\begin{figure}[b]
\begin{center}
\includegraphics[width=3.4in, clip=true]{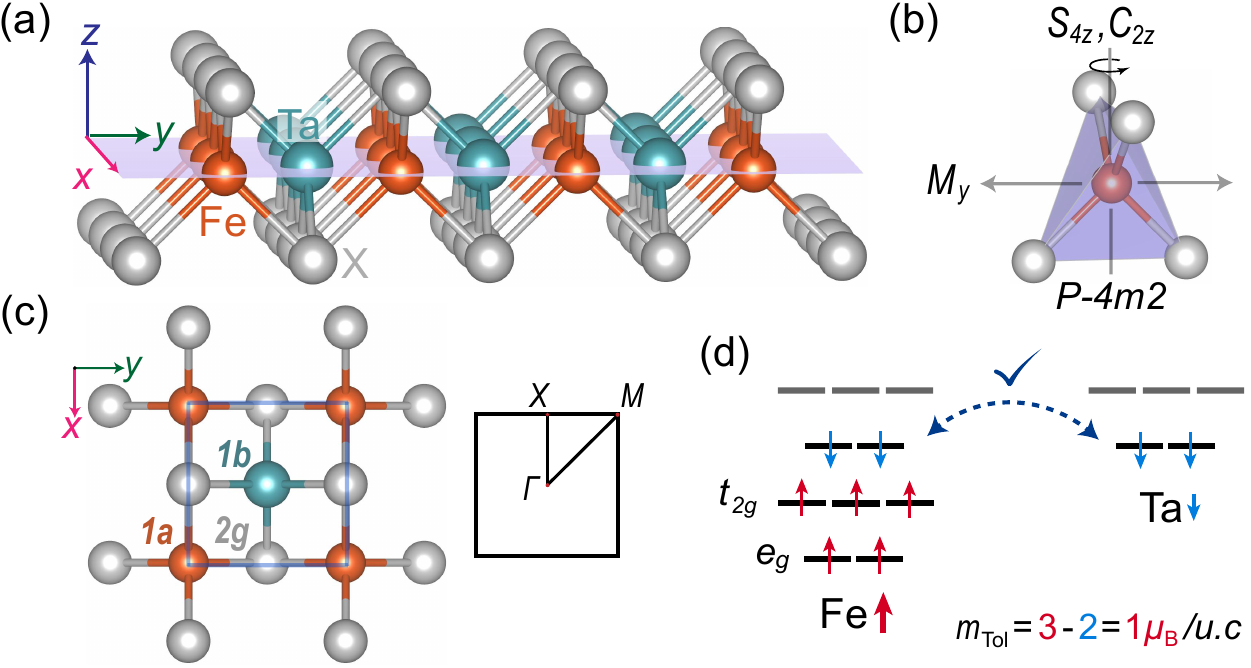}
\end{center}
\caption{(a),(c) Atomic structure of monolayer FeTa$X_2$ from side and top views. The Wyckoff positions $1a$, $1b$ and $2g$ are displayed (notation adopted from Bilbao crystallographic server~\cite{bilbao2,bilbao3,bilbao1,slager2017,vergniory2017,elcoro2017,bradlyn2017}). Brillouin zone is shown in (c). The 2D materials class has a collinear ferrimagnetic ground state along the $z$ direction with a total spin magnetic moment $1\mu_B$ per unit cell. (b) The key symmetry operations of $P$-$4m2$ include $S_{4z}$, $C_{2z}$ rotations and mirror symmetry $M_{y}~(M_x)$, where $S_{4z}\equiv\mathcal{I}C_{4z}$ and  $\mathcal{I}$ is inversion symmetry. (d) Schematic diagram of the crystal field splitting and $d$ orbital occupations for the Fe and Ta atoms in monolayer FeTa$X_2$. The AFM exchange coupling between Fe and Ta atoms can be interpreted as FM kinetic exchange coupling between the $e_g^{\downarrow 2}t_{2g}^{\downarrow 0}$ orbitals.}
\label{fig1}
\end{figure} 

\emph{Structure and magnetic properties---}The monolayer FeTa$X_2$ crystallizes in a tetragonal lattice structure with the space group $P$-$4m2$ (No.~115). As shown in Fig.~\ref{fig1}(a), each primitive cell includes three atomic layers—one FeTa and two $X_2$ layers—where each Fe or Ta atom is coordinated by four $X$ atoms forming a distorted edge-sharing tetrahedron. This structure can be viewed as a half-substituted derivative of Fe$X$ with space group $P4/nmm$, in which one of the two inequivalent Fe atoms in the unit cell is replaced by a Ta atom. Consequently, the inversion symmetry ($\mathcal{I}$) is broken, but the combined symmetry $S_{4z}\equiv\mathcal{I}C_{4z}$ remains preserved [Fig.~\ref{fig1}(b)]. The fully optimized lattice constants are listed in Table~\ref{tab1}. The dynamical and thermal stability of monolayer FeTa$X_2$ are confirmed by phonon calculations, formation energies and molecular dynamics simulations~\cite{supple}. 
In reality, FeCu$X_2$ with the same structure have been experimentally synthesized~\cite{kradinova1993novelCuFeX2,rivas1998mossbauerCuFeX2,llanos1995synthesisCuFeX2,ga2022anisotropicCuFeX2}. Meanwhile, their van der Waals layered structure implies that monolayers can be readily exfoliated from bulk crystals once synthesized.

The magnetic properties of monolayer FeTa$X_2$ are markedly altered by Ta substitution. First-principles calculations reveal that all compounds listed in Table~\ref{tab1} exhibit a robust collinear ferrimagnetic ground state with an out-of-plane easy axis, in contrast to monolayer FeTe, which is observed to have a stripe antiferromagnetic (AFM) ground state~\cite{enayat2014}. The total magnetic moment per unit cell is $1\mu_B$, where each Fe atom possesses sizable local moments ($\approx2.7\mu_B$), and Ta atom exhibits antiparallel moments ($\approx-1.4\mu_B$). The magnetic ordering is predominantly governed by the Fe atoms, while the moments of Ta atoms are induced via strong molecular fields generated by neighboring Fe atoms~\cite{supple}. This ferrimagnet effectively behaves as a ferromagnet, with the net magnetization aligned along the Fe moment direction. As a result, the system avoids a compensation temperature, consistent with N\'eel’s molecular field theory~\cite{goodenough1963,spaldin2012}.

To elucidate the origin of ferrimagnetism, we first analyze the orbital occupation of Fe atoms. In the tetrahedral crystal field, the Fe $3d$ orbitals split into a lower-energy $e_g$ doublet and a higher-energy $t_{2g}$ triplet [Fig.~\ref{fig1}(d)]. The Fe atoms adopt a high-spin $d^7$ configuration: $e_g^4t_{2g}^3=e_{g}^{\uparrow2}t_{2g}^{\uparrow3}e_g^{\downarrow2}t_{2g}^{\downarrow0}$, yielding a spin magnetic moment of $3\mu_B$ in accordance with Hund's rule, consistent with our DFT results. The spin-up channel is fully occupied, while the partially filled spin-down channel is at the Fermi level. Consequently, direct electron hopping between neighboring Fe sites is allowed only when their magnetic moments are aligned parallel, as enforced by the Pauli exclusion principle. In addition, the superexchange interaction mediated by the $X$ ligands (with the nearly $90^\circ$ Fe–$X$–Fe bond) further stabilize ferromagnetic (FM) coupling between the Fe atoms, according to the Goodenough–Kanamori–Anderson  rules~\cite{khomskii2004}. This particular $d^7$ configuration of Fe $3d$ orbitals in a tetrahedral crystal field leads to ultrastrong FM coupling between Fe atoms via both the FM kinetic exchange and superexchange mechanisms~\cite{sun2020,liy2020,yao2024}.

For the Ta atoms in FeTa$X_2$, the crystal fields generated by the surrounding Fe atoms and ligand $X$ atoms compete, resulting in a relatively small splitting between the lower-energy $e_g$ doublet and higher-energy $t_{2g}$ triplet of Ta $5d$ orbitals. As a result, each Ta atom adopts an $e_g^2t_{2g}^0$ configuration, yielding a local magnetic moment of approximately $2\mu_B$. The reduced crystal field splitting enhances the FM $e_g$-$t_{2g}$ kinetic exchange, which dominates over the AFM $e_g$-$e_{g}$ exchange between the neighboring Ta atoms~\cite{xuan2022,jiang2023}. Namely, the coupling between Ta atoms is FM. Moreover, direct electron hopping between Fe atoms in the $e_{g}^{\uparrow2}t_{2g}^{\uparrow3}e_g^{\downarrow2}t_{2g}^{\downarrow0}$ configuration and adjacent Ta atoms is energetically favorable when the latter adopt a $e^{\downarrow2}_gt^{\downarrow0}_{2g}$ configuration [Fig.~\ref{fig1}(d)]. This facilitates an effective AFM exchange coupling between the Fe and Ta sublattices.

Collectively, these exchange interactions stabilize a ferrimagnetic ground state in monolayer FeTa$X_2$, featuring strong FM coupling between Fe atoms and antiparallel alignment of the induced moments on Ta sites. The system exhibits exceptionally large MAE, exceeding that of known 2D magnets such as Fe$_3$GeTe$_2$~\cite{deng2018}. As listed in Table~\ref{tab1}, the large positive MAE indicates a robust out-of-plane easy axis. Notably, the predicted Curie temperature for monolayer FeTa$X_2$ significantly surpasses that of MnBi$_2$Te$_4$.

\begin{figure}[t]
\begin{center}
\includegraphics[width=3.4in, clip=true]{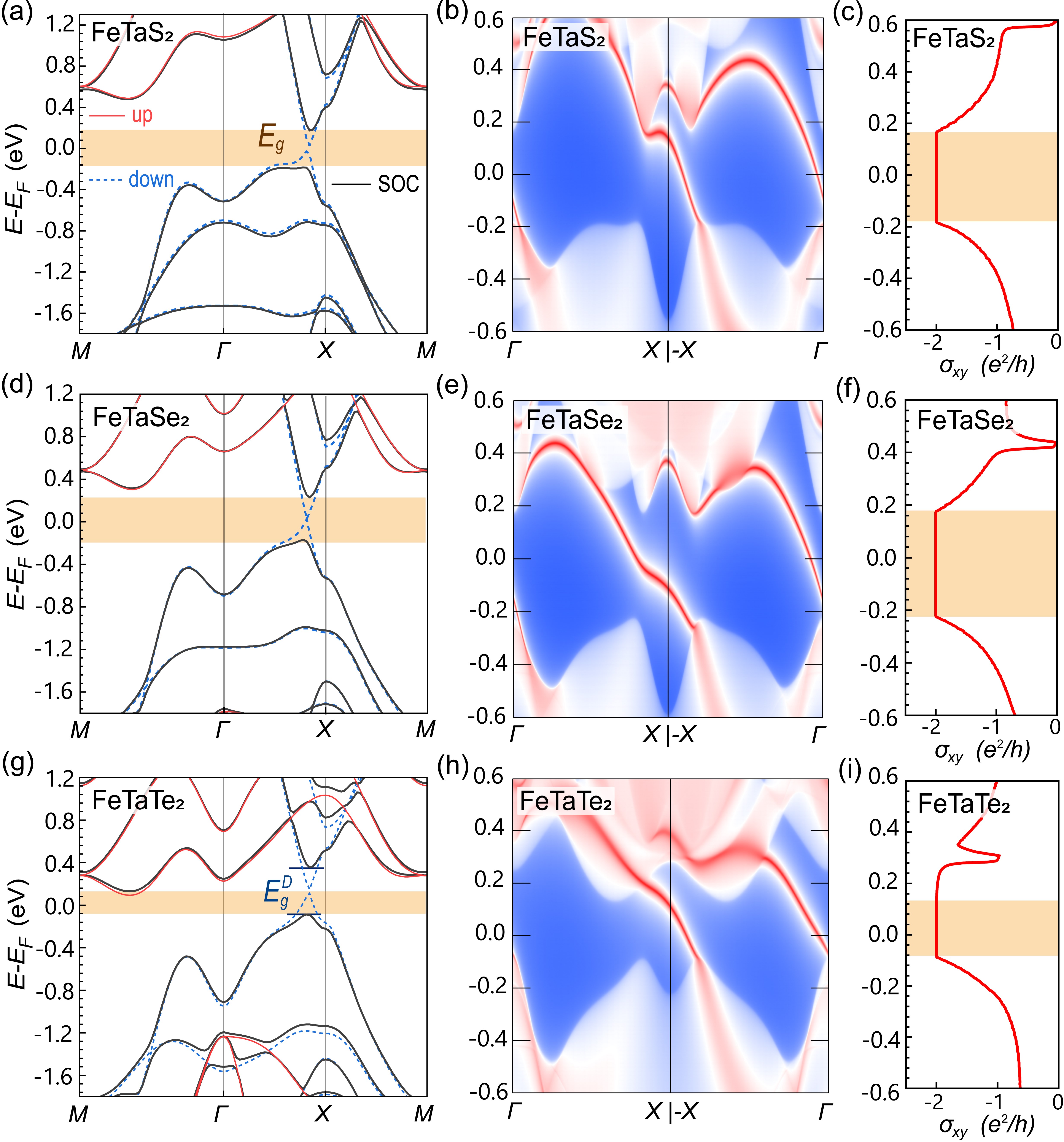}
\end{center}
\caption{Electronic structures and topological properties of monolayer FeTa$X_2$ ($X=$ S, Se, Te) by HSE method. (a)$-$(c) FeTaS$_2$, (d)$-$(f) FeTaSe$_2$, (g)$-$(i) FeTaTe$_2$, The band structure without (blue and red lines) and with (black line) SOC; topological edge states calculated along $x$ axis; and anomalous Hall conductance $\sigma_{xy}$ as a function of Fermi energy, respectively. The shaded regions in (a),(d),(g) and (c),(f),(i) denote the global topological gap $E_g$. (g) In FeTaTe$_2$, $E_g^{D}\approx449$~meV is the topologically nontrivial Dirac gap opened by SOC, with $E_g^{D}>E_g$. In FeTaS$_2$ and FeTaSe$_2$, $E_g^{D}=E_g$.}
\label{fig2}
\end{figure}

\emph{Electronic structures---} 
Fig.~\ref{fig2}(d) show the electronic structure of monolayer FeTaSe$_2$ with and without SOC. In the absence of SOC, four spin-polarized Dirac points emerge along the $\Gamma$-$\text{X}$(-$\text{Y}$) direction, which are protected by mirror symmetry $M_{y}$ ($M_x$). Upon inclusion of SOC, the out-of-plane ferrimagnetic ordering breaks both $M_{y}$ and $M_x$, opening a finite gap $E_g^D$ at each Dirac point. Each gapped Dirac cone contributes a quantized Berry phase of $-\pi$, yielding a total Berry phase of $-4\pi$ due to the fourfold valley degeneracy. This results in a QAH conductance of $\sigma_{xy} = -2e^2/h$ [Fig.~\ref{fig2}(f)], corresponding to a Chern number $\mathcal{C}=-2$. The nontrivial topology is further confirmed by the presence of two chiral edge modes traversing the bulk gap in the edge local density of states [Fig.~\ref{fig2}(e)]. The sign of $\mathcal{C}$ aligns with the direction of the Fe minority spins. The partial filling of high-Chern-number bands could lead to new topological states with exotic elementary excitations~\cite{barkeshli2012}. Substituting Se with isovalent chalcogen atoms S or Te generates similar band structures and preserves the overall band topology and Chern number, as illustrated in Figures~\ref{fig2}(a-c) and~\ref{fig2}(g-i). Notably, the Dirac gap $E_g^D$ increases from S to Te, reflecting the enhancement of SOC strength with heavier chalcogen elements.

\begin{figure}[b]
\begin{center}
\includegraphics[width=3.4in, clip=true]{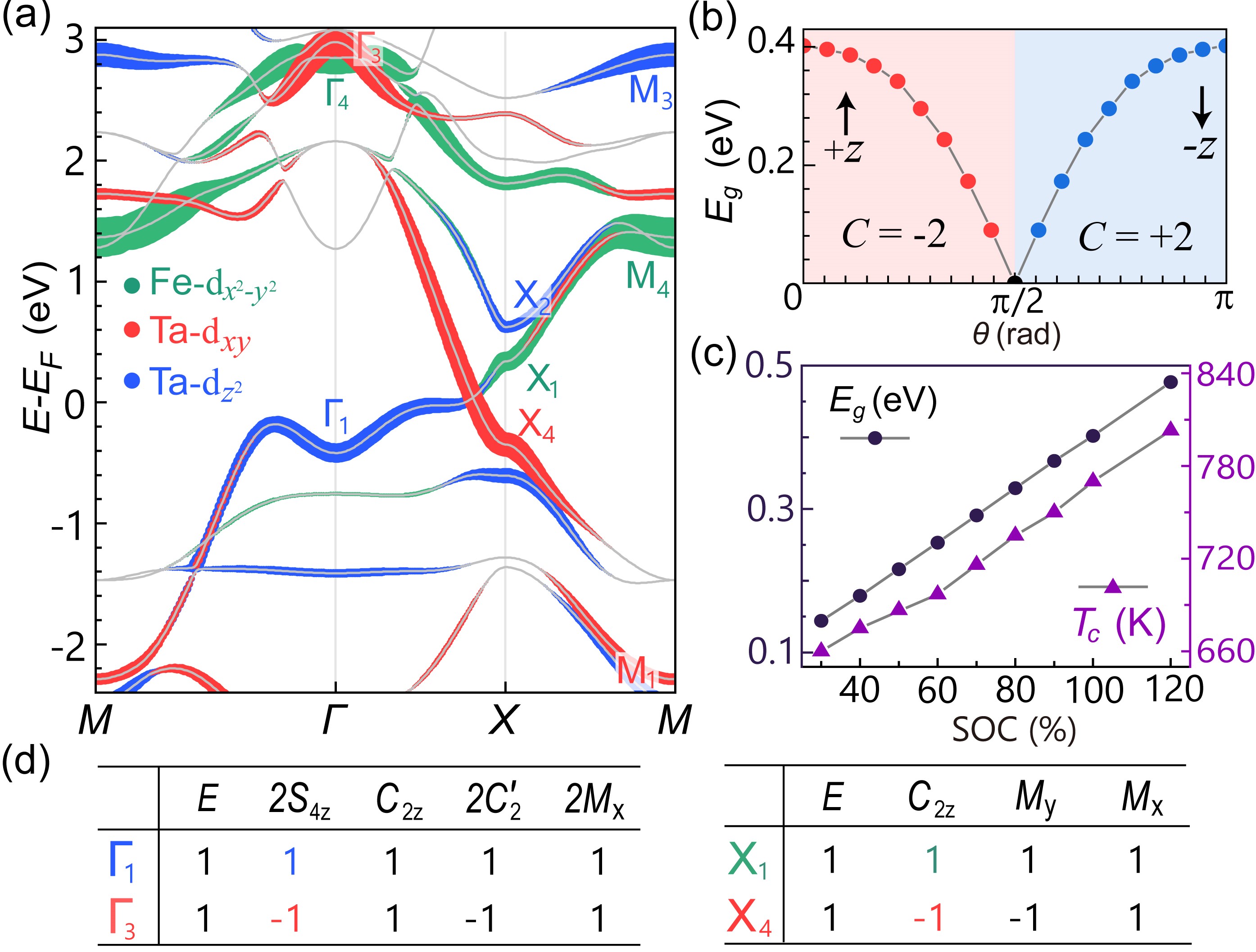}
\end{center}
\caption{(a) The $d$-orbitals projection band structures without SOC for spin down states of monolayer FeTaSe$_2$, only those bands which are related to band inversion are colored. The irreducible representation at high-symmetry points of the Brillouin zone boundary are displayed. (b) Dependence of the bulk gap and Chern number $\mathcal{C}$ on the spin orientation quantified by a polar angle $\theta$, where $\theta=0, \pi/2, \pi$ denote the $+z, +x, -z$ directions, respectively. (c) The evolution of the topological gap $E_g$ and Curie temperature $T_c$ with SOC strength. (d) Partial character table of little group at $\Gamma$ and X points. The rotation axis of $C_2^\prime$ symmetry is along $(\hat{x}-\hat{y})$ direction.  }
\label{fig3}
\end{figure}

The topological gap in FeTa$X_2$ exhibits a strong dependence on the spin orientation. As the magnetization direction rotates from $+z$ to $+x$, and subsequently to $-z$, the bulk gap closes and reopens, signaling two topological phase transitions—from $\mathcal{C}=-2$ to $\mathcal{C}=0$, and then to $\mathcal{C}=+2$  [Fig.~\ref{fig3}(b)]. Quantitatively, the gap follows an approximate relation $E_g\propto S \cos(\theta)\approx \langle S_z \rangle$,  indicating that the topological gap is predominantly governed by the out-of-plane spin component $\langle S_z \rangle$ of the Ta atoms via SOC. We point out that the large topological gap of the ground state still holds in the presence of magnon excitation, which is ensured by the large MAE~\cite{supple}. The significant thermal spin fluctuation will decrease $\langle S_z \rangle$ and the topological gap dramatically only when temperature is close to $T_c$.

In FeTa$X_2$, both the topology and magnetism originate from the $d$ orbitals of transition metal elements. To highlight their shared origin, Fig.~\ref{fig3}(c) presents the dependence of the Dirac gap and Curie temperature on the SOC strength in monolayer FeTaSe$_2$. As the SOC increases, both the topological gap and $T_c$ exhibit a clear positive correlation, indicating that SOC not only drives the nontrivial band topology but also enhances magnetic stability. This concurrent enhancement emphasizes a key advantage of $d$-orbital systems: the intrinsic coupling between magnetism and topology. Such synergy offers an effective strategy for realizing high-temperature QAH insulators with sizable topological gaps.

\emph{Model and origin of topology---}While the band structure near the Fermi level might suggest that the band topology arises from an $s$–$p$-type band inversion at the X point—supported by the opposite $C_{2z}$ eigenvalues of the irreducible representations X$_1$ and X$_4$ [Fig.~\ref{fig3}(d)] and their potential to yield a high Chern number ($\mathcal{C}=-2$) via valley multiplicity—a more detailed orbital analysis reveals otherwise. The $d$-orbital band projections and elementary band representation analysis in Fig.~\ref{fig3}(a) show that X$_1$ originates from the unoccupied Fe $d_{x^{2}-y^{2}}$ band, which does not participate in the band inversion. Instead, the nontrivial topology stems from the band inversion between the Ta $d^\downarrow_{z^{2}}$ and $d^\downarrow_{xy}$ orbitals at the $\Gamma$ point, involving an exchange of irreducible representations $\Gamma_3$ and $\Gamma_1$~\cite{supple}. Given the considerable depth of this inversion, its associated features are not readily apparent near the Fermi level.

\begin{figure}[b]
\begin{center}
\includegraphics[width=3.4in, clip=true]{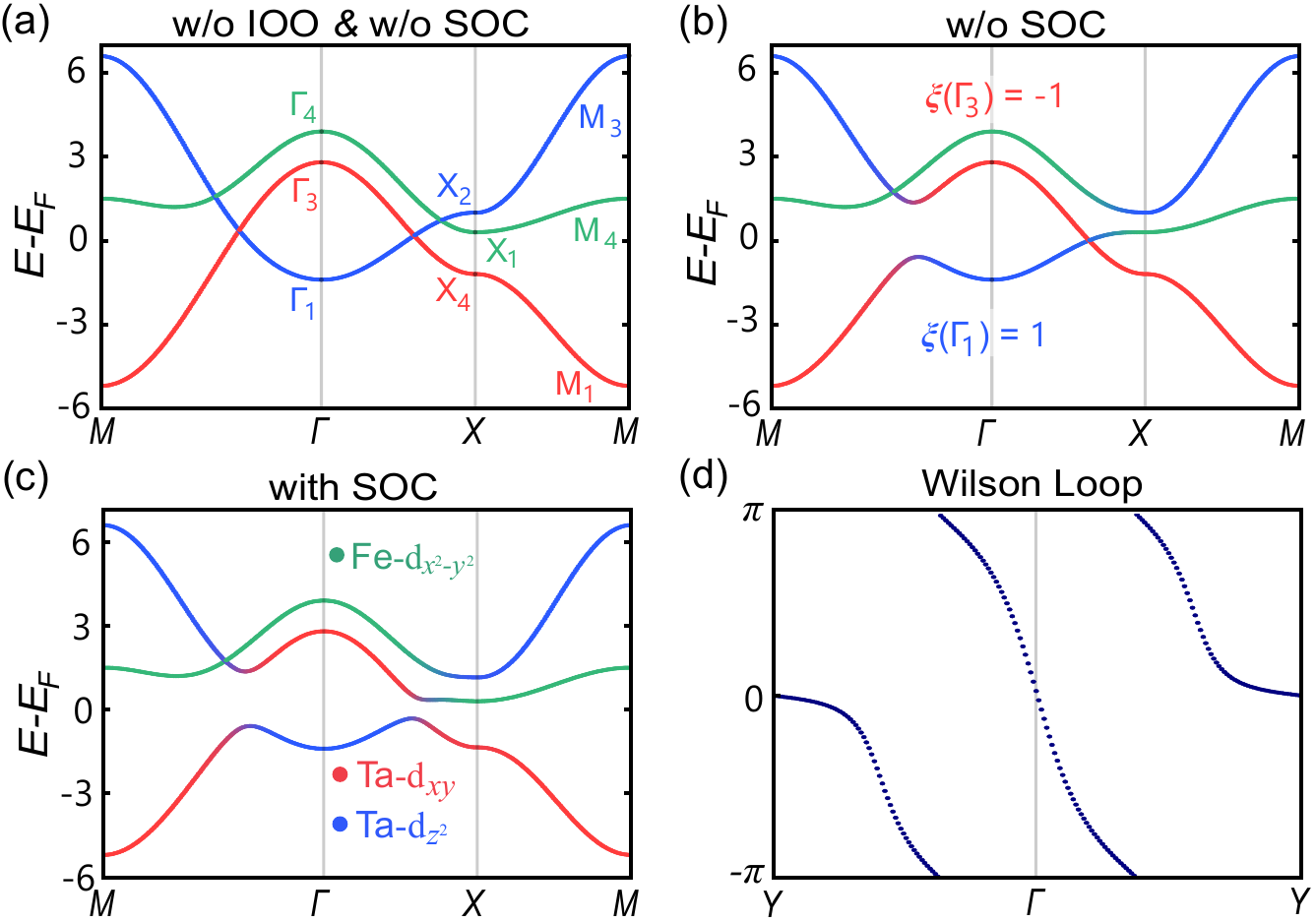}
\end{center}
\caption{The $d$-orbitals projection band structure of the tight-binding model (a) without inter-orbital overlap (IOO) and SOC, (b) with IOO but without SOC, (c) incorporating all relevant factors. The irreducible representation at high-symmetry points of the Brillouin zone boundary and the orbital compositions (color of the bands) are consistent with that in Fig.~\ref{fig3}(a). $\xi(\Gamma)$ is the eigenvalue of $S_{4z}$ at the $\Gamma$ point. (d) The Wilson loop for the lowest band in (c).}
\label{fig4}
\end{figure}

To elucidate the origin of the band topology, we construct a minimal tight-binding model that incorporates the Ta $d^{\downarrow}_{z^{2}},d^{\downarrow}_{xy}$, and Fe $d^{\downarrow}_{x^{2}-y^{2}}$ orbitals. This model, detailed in the Supplemental Material~\cite{supple}, includes the nearest- and next-nearest-neighbor hopping terms as well as SOC. As shown in Figures~\ref{fig4}(a) and~\ref{fig4}(b), it accurately reproduces the DFT band structure [Fig.~\ref{fig3}(a)] and the irreducible representations at high-symmetry points (listed in Table~\ref{tab2}). Importantly, while the model captures the apparent $s$–$p$-like features around the X point via inter-orbital hybridization, it demonstrates that the topological band structure is fundamentally driven by the $s$–$d$-type band inversion between the Ta $d^{\downarrow}_{z^{2}}$ and $d^{\downarrow}_{xy}$ orbitals at $\Gamma$ point.

We now analyze the origin of the large topological gap. The $s$–$d$ band inversion at the $\Gamma$ point can be described by a minimal $\mathbf{k}\cdot\mathbf{p}$ Hamiltonian,
\begin{equation}
  \label{eq1}
  \mathcal{H}_{\text{sd}}=
  \begin{pmatrix}
  -\frac{{k}_x^2+{k}_y^2}{2m_{1}}+\frac{\Delta}{2}  & \tilde{\lambda}i \left(k_x^2-k_y^2\right)+\tilde{t}\ k_xk_y \\ \tilde{\lambda}i \left(k_y^2-k_x^2\right)+\tilde{t}\ k_xk_y & \frac{{k}_x^2+{k}_y^2}{2m_{2}}-\frac{\Delta}{2}
  \end{pmatrix},
\end{equation}
where $m_{1}$ and $m_{2}$ are the effective mass of $d^\downarrow_{xy}$ and $d^{\downarrow}_{z^{2}}$ bands near the $\Gamma$ point, respectively. $\Delta$ characterizes the band inversion strength, typically several eV in FeTa$X_2$. The parameters $\tilde t$ and $\tilde \lambda$ are real: $\tilde t$ arises from inter-orbital overlap, while $\tilde \lambda$ originates from SOC. Due to the angular momentum difference $\delta\ell_z=2$ between the two orbitals, the inter-orbital couplings exhibit a quadratic $k$-dependence. This model leads to a topological gap $E_g \propto \tilde{\lambda}k^2$, which opens at finite $k$ away from the $\Gamma$ point due to orbital hybridization. In the case of FeTa$X_2$ with deep band inversion, where the relevant wave vectors lie near the X point, the gap is more accurately described by $E_g \propto \tilde{\lambda} \left(1-\cos k\right)$. This form highlights an enhancement of $E_g$ near the X point $k\to\pi$, with $1-\cos k \approx 2$,  in sharp contrast to systems with $s$–$p$ band inversion such as MnBi$_2$Te$_4$~\cite{zhang2019}, where $E_g \propto\tilde{\lambda}\sin k$. Combined with the strong SOC from the heavy Ta atoms, this deep $s$–$d$ inversion mechanism explains the large topological gaps summarized in Table~\ref{tab1}.

\begin{table}[t]
\caption{Partial elementary band representations without time-reversal symmetry for space group $P$-$4m2$. Fe and Ta take different Wyckoff positions in FeTa$X_2$, Fe at Wyckoff position $1a\equiv(0,0)$, Ta at Wyckoff position $1b\equiv(1/2,1/2)$.} 
\begin{center}\label{tab2}
\renewcommand{\arraystretch}{1.4}
\begin{tabular*}{3.4in}
{@{\extracolsep{\fill}}cccc}
\hline
\hline
  & $\Gamma$ & $X$ & $M$\\ 
\hline
  $d_{z^2}$@Ta & $\Gamma_1(1)$ & $X_2(1)$ & $M_3(1)$\\
  $d_{xy}$@Ta & $\Gamma_3(1)$ & $X_4(1)$ & $M_1(1)$\\
  $d_{x^2-y^2}$@Fe & $\Gamma_4(1)$ & $X_1(1) $ & $M_4(1)$\\
\hline
\hline
\end{tabular*}
\end{center}
\end{table}

\emph{Discussions---}The high-temperature QAH states in FeTa$X_2$ arise from cooperative mechanisms of topology and magnetism entirely within $d$ orbitals. The nontrivial band topology is driven by SOC-induced inversion of Ta $5d$ orbitals, while robust magnetism originates from Fe $3d$ electrons. The Fe–Ta $d$-orbital hybridization is substantial~\cite{supple}. In sharp contrast, MnBi$_2$Te$_4$ derives its topology from Bi/Te $p$ orbitals at the Fermi level, while magnetism comes from Mn $3d$ orbitals located far from the gap. The weak $p$–$d$ hybridization between Mn and Te yields relatively weak exchange interactions, which not only suppress the critical temperature but also limit the size of the topological gap. This cooperative interplay of magnetism and topology arising from $d$-orbital physics highlights a broadly applicable design principle for realizing QAH phases with enhanced gaps and elevated critical temperatures.

FeTa$X_2$ constitutes the first example of a QAH insulator in a stoichiometric ferrimagnet. Owing to their inequivalent magnetic sublattices, ferrimagnets can potentially offer the combined advantages of both ferromagnets and antiferromagnets~\cite{kim2022}: net magnetization is readily controllable via external fields, while spin dynamics remain fast, akin to AFM systems. The coexistence of topology and ferrimagnetism could open a promising new research direction in the interdisciplinary fields of topological materials and spintronics. It is intriguing to find compensated ferrimagnetic QAH insulators, in which the topological quantization efffect and dissipationless spintronic devices can be realized without any net magnetization and be controlled via fast magnetization reversal~\cite{finley2020}.

FeTa$X_2$ also bears an intrinsic connection to superconductivity. Its lattice structure closely matches that of FeSe and related superconductors, enabling the fabrication of FeTa$X_2$/Fe$X$ heterostructures. The resulting superconducting proximity effect, combined with reduced net magnetization, provides a viable platform for realizing chiral topological superconductivity and Majorana fermions~\cite{qi2010b,wang2015c,lian2018b}. Moreover, drawing inspiration from Li$_x$FeSe—where increasing lithium concentration drives a crossover from a nonmagnetic superconductor to a superconducting ferromagnet~\cite{ying2018,kim2023,hu2025Li}—one may tune the substitution ratio in Fe$_{2-x}$Ta$_xX_2$ ($0 \leq x \leq 1$). Such tuning could yield a rich phase diagram that may evolve from a nonmagnetic superconductor to a time-reversal-breaking superconducting magnet, and ultimately to a QAH insulator.

In summary, we identify a family of large-gap, high-temperature QAH insulators with high Chern number $\mathcal{C}=-2$, in which magnetism and topology emerge cooperatively from $d$ orbital electrons. In contrast to earlier generations of magnetic TIs—such as magnetically doped TI and MnBi$_2$Te$_4$—the FeTa$X_2$ compounds uniquely combine structural simplicity, intrinsic magnetism, and nontrivial topology. We anticipate that these materials will serve as a next-generation platform for exploring topological quantum matter and advancing spintronic applications.

\begin{acknowledgments}
\emph{Acknowledgments---}We thank J. Gao, T. Yu, J. Xiao, and Y. Zhang for valuable discussions. This work is supported by the Natural Science Foundation of China through Grants No.~12350404 and No.~12174066, the Innovation Program for Quantum Science and Technology through Grant No.~2021ZD0302600, the Science and Technology Commission of Shanghai Municipality under Grants No.~23JC1400600, No.~24LZ1400100 and No.~2019SHZDZX01, and is sponsored by the ``Shuguang Program'' supported by the Shanghai Education Development Foundation and Shanghai Municipal Education Commission. Y.J. acknowledges additional support from the China Postdoctoral Science Foundation under Grants No. GZC20240302 and No. 2024M760488. 

Y.J. and H.W. contributed equally to this work.
\end{acknowledgments}

\end{document}